\documentstyle[12pt]{article}
\begin{document}
\thispagestyle{empty}

\newcommand{\p}[1]{(\ref{#1})}
\newcommand{\be}{\begin{equation}}
\newcommand{\ee}{\end{equation}}
\newcommand{\sect}[1]{\setcounter{equation}{0}\section{#1}}

\newcommand{\vs}[1]{\rule[- #1 mm]{0mm}{#1 mm}}
\newcommand{\hs}[1]{\hspace{#1mm}}
\newcommand{\mb}[1]{\hs{5}\mbox{#1}\hs{5}}
\newcommand{\Db}{{\overline D}}
\newcommand{\bea}{\begin{eqnarray}}
\newcommand{\eea}{\end{eqnarray}}
\newcommand{\wt}[1]{\widetilde{#1}}
\newcommand{\und}[1]{\underline{#1}}
\newcommand{\ov}[1]{\overline{#1}}
\newcommand{\sm}[2]{\frac{\mbox{\footnotesize #1}\vs{-2}}
		   {\vs{-2}\mbox{\footnotesize #2}}}
\newcommand{\prt}{\partial}
\newcommand{\eps}{\epsilon}

\newcommand{\R}{\mbox{\rule{0.2mm}{2.8mm}\hspace{-1.5mm} R}}
\newcommand{\Z}{Z\hspace{-2mm}Z}

\newcommand{\cd}{{\cal D}}
\newcommand{\cg}{{\cal G}}
\newcommand{\ck}{{\cal K}}
\newcommand{\cw}{{\cal W}}

\newcommand{\vj}{\vec{J}}
\newcommand{\vl}{\vec{\lambda}}
\newcommand{\vz}{\vec{\sigma}}
\newcommand{\vt}{\vec{\tau}}
\newcommand{\vw}{\vec{W}}
\newcommand{\poiss}{\stackrel{\otimes}{,}}

\def\l#1#2{\raisebox{.2ex}{$\displaystyle
  \mathop{#1}^{{\scriptstyle #2}\rightarrow}$}}
\def\r#1#2{\raisebox{.2ex}{$\displaystyle
 \mathop{#1}^{\leftarrow {\scriptstyle #2}}$}}


\newcommand{\NP}[1]{Nucl.\ Phys.\ {\bf #1}}
\newcommand{\PL}[1]{Phys.\ Lett.\ {\bf #1}}
\newcommand{\NC}[1]{Nuovo Cimento {\bf #1}}
\newcommand{\CMP}[1]{Comm.\ Math.\ Phys.\ {\bf #1}}
\newcommand{\PR}[1]{Phys.\ Rev.\ {\bf #1}}
\newcommand{\PRL}[1]{Phys.\ Rev.\ Lett.\ {\bf #1}}
\newcommand{\MPL}[1]{Mod.\ Phys.\ Lett.\ {\bf #1}}
\newcommand{\BLMS}[1]{Bull.\ London Math.\ Soc.\ {\bf #1}}
\newcommand{\IJMP}[1]{Int.\ Jour.\ of\ Mod.\ Phys.\ {\bf #1}}
\newcommand{\JMP}[1]{Jour.\ of\ Math.\ Phys.\ {\bf #1}}
\newcommand{\LMP}[1]{Lett.\ in\ Math.\ Phys.\ {\bf #1}}

\renewcommand{\thefootnote}{\fnsymbol{footnote}}
\newpage
\setcounter{page}{0}
\pagestyle{empty}
\begin{flushright}
{April 1997}\\
{SISSA 52/97/EP}\\
{hep-th/9704130}
\end{flushright}
\vs{8}
\begin{center}
{\LARGE {\bf The Hamiltonian structure of the}}\\[0.6cm]
{\LARGE {\bf $N=2$ supersymmetric GNLS hierarchy}}\\[1cm]

\vs{8}

{\large L. Bonora$^{a,1}$ and A. Sorin$^{b,2}$}
{}~\\
\quad \\
{\em ~$~^{(a)}$ International School for Advanced Studies (SISSA/ISAS),}\\
{\em Via Beirut 2, 34014 Trieste, Italy}\\
{\em INFN, Sezione di Trieste}\\
{\em {~$~^{(b)}$ Bogoliubov Laboratory of Theoretical Physics, JINR,}}\\
{\em 141980 Dubna, Moscow Region, Russia}~\quad\\

\end{center}
\vs{8}

\centerline{ {\bf Abstract}}
\vs{4}

The first two Hamiltonian structures and the recursion operator connecting
all evolution systems and Hamiltonian structures of the $N=2$
supersymmetric $(n,m)$-GNLS hierarchy are constructed in terms of $N=2$
superfields in two different superfield bases with local evolution
equations. Their bosonic limits are studied in detail. New local and
nonlocal bosonic and fermionic integrals both for the $N=2$ supersymmetric 
$(n,m)$-GNLS hierarchy and its bosonic counterparts are derived. As an 
example, in the n=1, m=1 case, the algebra and the symmetry transformations
for some of them are worked out and a rich N=4 supersymmetry structure 
is uncovered.

\vfill
{\em E-Mail:\\
1) bonora@sissa.it\\
2) sorin@thsun1.jinr.dubna.su }
\newpage

\pagestyle{plain}
\renewcommand{\thefootnote}{\arabic{footnote}}
\setcounter{footnote}{0}

\noindent{\bf 1. Introduction.}
Recently there has been an intense research activity on N=2 supersymmetric
integrable hierarchy. Basically this stems from
the interest spurred by the ordinary integrable hierarchies and their relation 
to the 2D gravity and N=2 supersymmetric models representing superstring 
vacua. Although an analysis of this connection for the N=2 supersymmetric
integrable hierarchies is still lacking, it is nevertheless a fact that the
latter hierarchies have extremely rich and interesting structures. In \cite{bks}
a large class of such hierarchies was introduced: the $N=2$ supersymmetric
$(n,m)$ Generalized Nonlinear Schr\"{o}dinger (GNLS) hierarchies. They were
subsequently studied in a number of other papers, \cite{ara,dg,ik,dls,s1}.

The goal of the present letter is to fill a gap in our knowledge of these
hierarchies by analyzing their Hamiltonian structures. In particular we
produce here the first two Hamiltonian structures and the relevant recursion 
operators, as well as related local and non--local conserved charges, in two
different superfield representations which possess local flow equations.

Let us start with a short summary of the main facts concerning
the $N=2$ supersymmetric $(n,m)$-GNLS hierarchy \cite{bks,s1}
which will be useful in what follows.

The Lax operator of the $N=2$ supersymmetric $(n,m)$-GNLS
hierarchy has the following form\footnote{Summation over
repeated indices is understood and the square brackets mean that the relevant
operators act only on superfields inside the brackets.}
\begin{eqnarray}
L= \partial - \frac{1}{2}(F_a {\overline F}_a +
F_a {\overline D}\partial^{-1} \left[ D {\overline F}_a\right]), \quad
[D,L]=0,
\label{suplax}
\end{eqnarray}
where $F_a(Z)$ and ${\overline F}_a(Z)$ ($a,b=1,\ldots , n+m$) are
chiral and antichiral $N=2$ superfields
\begin{eqnarray}
D F_a(Z)=0, \quad {\overline D}~{\overline F}_a(Z) = 0,
\label{chiral}
\end{eqnarray}
respectively. They are bosonic for $a=1,\ldots ,n$ and fermionic for
$a=n+1,\ldots , n+m$, i.e., $F_aF_b=(-1)^{d_ad_b}F_bF_a$, where
$d_a=1$ $(d_a=0)$ is the Grassman parity for fermionic (bosonic) superfields;
$Z=(z,\theta,\overline\theta)$ is a coordinate of the $N=2$ superspace, $dZ
\equiv dz d \theta d \overline\theta$ and $D,{\overline D}$ are the $N=2$
supersymmetric fermionic covariant derivatives
\begin{eqnarray}
D=\frac{\partial}{\partial\theta}
 -\frac{1}{2}\overline\theta\frac{\partial}{\partial z}, \quad
{\overline D}=\frac{\partial}{\partial\overline\theta}
 -\frac{1}{2}\theta\frac{\partial}{\partial z}, \quad
D^{2}={\overline D}^{2}=0, \quad
\left\{ D,{\overline D} \right\}= -\frac{\partial}{\partial z}
\equiv -{\partial}.
\label{DD}
\end{eqnarray}
For $p=0,1,2,.. $, the Lax operator $L$ provides the consistent flows
\begin{eqnarray}
{\textstyle{\partial\over\partial t_p}}L =[ (L^p)_{\geq 1} , L].
\label{laxfl1}
\end{eqnarray}
An infinite number of Hamiltonians can be obtained as follows:
\begin{eqnarray}
{H}_p =\int d Z {\cal H}_p, \quad  {\cal H}_p \equiv (L^p)_{0},
\label{res}
\end{eqnarray}
where the subscripts $\geq 1$ and $0$ mean the sum of the purely
derivative terms and the constant part of the operator, 
respectively\footnote{An alternative Lax representation 
of the $N=2$ supersymmetric $(n,m)$-GNLS hierarchy was proposed in
\cite{dg}. Its relation to our Lax representation is not completely
clear to us.}. The evolution equations \p{laxfl1} for the
superfields $F_a$ and $\overline F_a$ are local,
\begin{eqnarray}
{\textstyle{\partial\over\partial t_p}}F_a = ((L^p)_{\geq 1} F_a)_0, \quad
{\textstyle{\partial\over\partial t_p}}\overline F_a =(-1)^{p+1}
((L^{*~p})_{\geq 1} \overline F_a)_0,
\label{n1}
\end{eqnarray}
and they admit the complex structure
\begin{eqnarray}
F_a^{*}= (-i)^{d_a-1} {\overline F}_a, \quad
{\overline F_a^{*}}=(-i)^{d_a-1} F_a, \quad
{\theta}^{*}={\overline {\theta}}, \quad
{\overline {\theta}^{*}}={\theta}, \quad
t^{*}_p= (-1)^{p+1} t_p, \quad
z^{*}= z,
\label{conj}
\end{eqnarray}
where $i$ is the imaginary unity, and $L^*$ is the complex-conjugate Lax
operator
\begin{eqnarray}
L^*= \partial + \frac{1}{2}(F_a {\overline F}_a + (-1)^{d_a}
\overline F_a D\partial^{-1} \left[ {\overline D} F_a\right]), \quad
[~\overline D, L^*]=0,
\label{suplaxconj}
\end{eqnarray}
which also provides consistent flows.

The first three flows from \p{n1} and the first three nontrivial
Hamiltonian densities from \p{res} read as:
\begin{eqnarray}
&&{\textstyle{\partial\over\partial t_0}} F_a = F_a,\quad
{\textstyle{\partial\over \partial t_0}} {\overline F}_a =
-{\overline F}_a; \quad
{\textstyle{\partial\over\partial t_1}} F_a =F_a~', \quad
{\textstyle{\partial\over \partial t_1}} {\overline F}_a =
{\overline F}_a~';\nonumber\\
&&{\textstyle{\partial\over\partial t_2}} F_a =
F_a~'' +  {D}(F_b {\overline F}_b~{\overline D} F_a), \quad
{\textstyle{\partial\over \partial t_2}} {\overline F}_a =
-{\overline F}_a~'' + {\overline D}
(F_b {\overline F}_b {D}{\overline F}_a),
\label{gnls}
\end{eqnarray}
\begin{eqnarray}
&&{\cal H}_1 = -\frac{1}{2} F_a {\overline F}_a, \quad
{\cal H}_2 = \frac{1}{2}( F_a {\overline F}_a~' +
\frac{1}{4} (F_a {\overline F}_a)^2 ), \nonumber\\
&& {\cal H}_3 = -\frac{1}{2} ( F_a {\overline F}_a~'' -
\frac{1}{2} \left[{\overline D} F_a {\overline F}_a\right]
\left[D F_b {\overline F}_b\right] +
F_a {\overline F}_a~' F_b {\overline F}_b +
\frac{1}{12} (F_a {\overline F}_a)^3 ),
\label{i1}
\end{eqnarray}
respectively, where $'$ means the derivative with respect to z.
The second flow from the set \p{gnls} forms the $N=2$ supersymmetric
$(n,m)$-GNLS equations.

It is useful, as will be clear in a moment, to introduce an alternative 
superfield basis by means of
$\{J(Z), {\overline \Phi}_j(Z), \Phi_j(Z),
j=1,\ldots , l-1, l+1, \ldots , n, \ldots , n+m\}$,
\begin{eqnarray}
J=\frac{1}{2}(\frac{1}{2} F_a {{\overline F}_a} - (\ln F_l)~'~), \quad
{\overline \Phi}_j=\frac{1}{\sqrt{2}} \overline D ({F_l}^{-1}F_{j}),\quad
\Phi_j = \frac{1}{\sqrt{2}} D (F_l{\overline F}_{j}),
\label{newbasis}
\end{eqnarray}
where the index $l$ is an arbitrary fixed index belonging to the range
$1\leq l \leq n$. A different choice of $l$ leads to different bases, 
but in the LHS of \p{newbasis} for simplicity we drop the 
symbol $l$. The flows \p{gnls} and Hamiltonian densities \p{i1} now 
become
\begin{eqnarray}
&&{\textstyle{\partial\over\partial t_0}}J=
{\textstyle{\partial\over\partial t_0}}{\overline \Phi}_i=
{\textstyle{\partial\over\partial t_0}}\Phi_i=0; \quad
{\textstyle{\partial\over\partial t_1}}J=J~',\quad
{\textstyle{\partial\over\partial t_1}}{\overline \Phi}_i=
{\overline \Phi}_i~', \quad
{\textstyle{\partial\over\partial t_1}}\Phi_i=\Phi_i~'; \nonumber\\
&&{\textstyle{\partial\over\partial t_2}}{J} =
(-[D,{\overline D}~] J - 2 J^2+ {\Phi_j}{\overline \Phi}_j)~', \nonumber\\
&&{\textstyle{\partial\over\partial t_2}}{\Phi_j} =
-{\Phi_j}~''+ 4D\overline D(J \Phi_j), \quad
{\textstyle{\partial\over\partial t_2}}{\overline \Phi}_j =
{\overline \Phi}_j~''+ 4\overline D D (J \overline \Phi_j),
\label{newgnls}
\end{eqnarray}
\begin{eqnarray}
{\cal H}_1 = -2 J, \quad
{\cal H}_2 =  2 J^2 -\Phi_j {\overline \Phi}_j, \quad
{\cal H}_3 =  \Phi_j~'~{\overline \Phi}_j +
4 J \Phi_j {\overline \Phi}_j - 4 {\overline D}J DJ - \frac{8}{3} J^3,
\label{i2}
\end{eqnarray}
respectively\footnote{Let us recall that Hamiltonian densities
are defined up to terms which are fermionic or bosonic total
derivatives of arbitrary nonsingular, local functions of the superfields.}.
In addition to the first complex structure \p{conj} hidden in this basis,
they admit an extra, second complex structure \begin{eqnarray}
{\Phi}_j^{*}= (-i)^{d_j} {\overline \Phi}_j,\quad
{\overline \Phi}_j^{*}=(-i)^{d_j}{\Phi}_j, \quad J^{*} = -J, \quad
{\theta}^{*}={\overline {\theta}}, \quad
{\overline {\theta}^{*}}={\theta}, \quad
t^{*}_p= (-1)^{p+1} t_p, \quad z^{*}= z, ~
\label{nnconj}
\end{eqnarray}
which is manifest in this basis, but it is hidden in the former one.
Here, $d_i$ is the Grassman parity of the superfields
$\Phi_i$ and ${\overline \Phi}_i$. We call the basis
\p{newbasis} a KdV-basis, reflecting the fact that for $n=1, m=0$, eqs.
\p{newgnls} form the flows of the $N=2$ supersymmetric $a=4$ KdV hierarchy
\cite{lm}. In the KdV-basis, the $N=2$ supersymmetric $(n,m)$-GNLS
hierarchy of integrable equations, together with its Hamiltonians, can be
produced using formulas \p{laxfl1}, \p{res}, where the Lax operator $L$
\p{suplax} is replaced by the gauge related Lax operator
\begin{eqnarray}
L^{KdV} = F_l^{-1} L F_l \equiv
\partial - 2J-2 {\overline D}\partial^{-1}
\left[ D (J -\frac{1}{2}{\Phi}_j \partial^{-1} {\overline {\Phi}}_j)\right]
+ \left[ D \partial^{-1} {\overline {\Phi}}_j\right]
{\overline D}\partial^{-1} {\Phi}_j.
\label{kdvsuplax}
\end{eqnarray}
For general values of the discrete parameters $n$ and
$m$, the Lax representation of the hierarchy corresponding to eqs.
\p{newgnls} was proposed in \cite{ik}, and its relationships to the
$N=2$ supersymmetric $(n,m)$-GNLS hierarchy was established in \cite{s1}.
In addition to the transformations \p{newbasis}, relating egs.\p{newgnls} to
\p{gnls}, there are other transformations (for details, see \cite{s1});
however, for our purpose here, it will be enough to consider only
these transformations.

{}~

\noindent{\bf 2. Hamiltonian structure of the $N=2$ super
$(n,m)$-GNLS hierarchy.}
A bi-Hamiltonian system of evolution equations can be represented in the
following general form:
\begin{eqnarray}
&& {\textstyle{\partial\over\partial t_p}}
\left(\begin{array}{cc} F_a\\{\overline F_a} \end{array}\right) =
(J_1)_{ab}\left(\begin{array}{cc} {\delta}/{\delta F_b} \\
{\delta}/{\delta {\overline F}_b} \end{array}\right) H_{p+1}=
(J_2)_{ab}\left(\begin{array}{cc} {\delta}/{\delta F_b} \\
{\delta}/{\delta {\overline F}_b} \end{array}\right) H_{p}, \nonumber\\
&&(J^{(-1)}_1)_{ab} {\textstyle{\partial\over\partial t_p}}
\left(\begin{array}{cc} F_b\\{\overline F}_b \end{array}\right) =
\left(\begin{array}{cc} {\delta}/{\delta F_a} \\
{\delta}/{\delta {\overline F}_a} \end{array}\right) H_{p+1},
\label{hameq}
\end{eqnarray}
where $J_1$ and $J_2$ are the first and second Hamiltonian structures. Here
we have introduced also the matrix $J^{(-1)}_1$ defined by the relations:
\begin{eqnarray}
J_1 J^{(-1)}_1= \Pi  ,\quad J^{(-1)}_1 J_1  = {\overline {\Pi}} \quad
\iff \{ J_1 , J^{(-1)}_1 \} = I,
\label{propstarj}
\end{eqnarray}
where $\Pi$ (${\overline {\Pi}}$)
\begin{eqnarray}
\Pi & \equiv & - \left(\begin{array}{cc} D\overline D \partial^{-1}
{\delta}_{ab}, & 0 \\
0, & \overline D D \partial^{-1}{\delta}_{ab}
\end{array}\right), \quad
{\overline {\Pi}} \equiv - \left(\begin{array}{cc} {\overline D}
D \partial^{-1}{\delta}_{ab}, & 0 \\
0, & D{\overline D}  \partial^{-1}{\delta}_{ab}
\end{array}\right), \nonumber\\
&& \Pi \Pi =\Pi, \quad {\overline \Pi}~ {\overline \Pi}=\overline \Pi,
\quad \Pi \overline \Pi=\overline \Pi \Pi=0, \quad \Pi + \overline \Pi =I
\label{pi}
\end{eqnarray}
is the matrix that projects the up and down elements of a column on
the chiral (antichiral) and antichiral (chiral) subspaces, respectively.
In terms of the Hamiltonian structure $J_p$, the $N=2$ supersymmetric
Poisson brackets algebra of the superfields $F_a$ and ${\overline F}_a$
are given by the formula:
\begin{eqnarray}
\{\left(\begin{array}{cc}
F_a(Z_1)\\{\overline F_a(Z_1)}\end{array}\right)
\stackrel{\otimes}{,}
\left(\begin{array}{cc} F_b(Z_2),{\overline F_b(Z_2)}\end{array}\right)
\}_p=(J_p)_{ab}(Z_1){\delta}^{N=2}(Z_1-Z_2),
\label{palg}
\end{eqnarray}
where ${\delta}^{N=2}(Z) \equiv \theta {\overline \theta} {\delta}(z)$ is
the delta function in $N=2$ superspace and the notation `$\otimes$' stands
for the tensor product. In addition to the Jacobi identities and symmetry
properties respecting the statistics of the superfields, $J_p$ should also
satisfy the chiral consistency conditions
\begin{eqnarray}
J_p \Pi={\overline \Pi} J_p=0, \quad J_p {\overline \Pi}= \Pi J_p=J_p,
\label{cons}
\end{eqnarray}
which shows that all the Hamiltonian structures are represented by degenerate
matrices. This is the peculiarity of a manifest $N=2$ superinvariant
description of the $N=2$ supersymmetric $(n,m)$-GNLS hierarchy in terms
of $N=2$ superfields, which has no analogue in the description in terms
of $N=1$ superfields or components. One should stress that this is not
a pathology of the Hamiltonian structures, but a peculiarity of the $N=2$ 
superfield description, which can be easily dealt with.

Using the first three flows of the $N=2$ supersymmetric $(n,m)$-GNLS
hierarchy and Hamiltonians with densities \p{i1}, we have found its first
two Hamiltonian structures. The explicit expressions
for them as well as for the recursion operator of the hierarchy are
presented below. We postpone the discussion of their consistency
(the Jacobi identities, the compatibility of the Hamiltonian
structures and the hereditarity \cite{ff} of the recursion operator)
till the end of the next section, where we construct their explicit 
expressions in the KdV-basis, which are more suitable for this purpose.

In spite of the very complicated form of the first Hamiltonian structure
$J_1$, the expression for its inverse matrix $J^{(-1)}_1$ is quite simple
and looks like\footnote{Hereafter, it is understood that the derivatives 
$\partial$, ${\overline D}$ and $D$ appearing in the Hamiltonian structures,
are to be considered as operators that act on whatever is on their right.}
\begin{eqnarray}
(J^{(-1)}_1)_{ab} = \frac{1}{4} {\overline \Pi} \left(\begin{array}{cc}
(-1)^{d_b}{\overline F}_a \partial^{-1} {\overline F}_b,
& {\overline F}_a \partial^{-1} F_b + 2 {\delta}_{ab} \\
(-1)^{d_a}((-1)^{d_b}F_a \partial^{-1} {\overline F}_b-2{\delta}_{ab}),
& (-1)^{d_a}F_a \partial^{-1} F_b
\end{array}\right) \Pi.
\label{starj}
\end{eqnarray}
Actually, in what follows, we need only the explicit expression for the
matrix $J^{(-1)}_1$ and due to the very complicated form of $J_1$, we do
not present it here.

The second Hamiltonian structure has the following form:
\begin{eqnarray}
&& (J_2)_{ab}= \left(\begin{array}{cc} (J_{11})_{ab}, & (J_{12})_{ab} \\
(J_{21})_{ab} &  (J_{22})_{ab}
\end{array}\right),  \nonumber\\
&& (J_{11})_{ab}=(-1)^{d_ad_b} F_b D{\overline D} \partial^{-1} F_a-
F_a D{\overline D} \partial^{-1} F_b, \nonumber\\
&& (J_{12})_{ab}=(-1)^{d_b}(2D \overline D -F_cD
{\overline D}\partial^{-1}{\overline F}_c){\delta}_{ab}+
F_a D{\overline D} \partial^{-1} {\overline F}_b, \nonumber\\
&& (J_{21})_{ab}=(2{\overline D} D +(-1)^{d_c}{\overline F}_c~{\overline
D}D\partial^{-1} F_c){\delta}_{ab} -
{\overline F}_a~{\overline D}D\partial^{-1}F_b, \nonumber\\
&& (J_{22})_{ab}= {\overline F}_a~{\overline D}D{\partial^{-1}}
{\overline F}_b - (-1)^{d_ad_b}{\overline F}_b~{\overline D}D
{\partial^{-1}}{\overline F}_a.
\label{hamstr2}
\end{eqnarray}

Knowledge of the first and second Hamiltonian structures allows us to
construct the recursion operator $R_{ab}$ of the $N=2$ supersymmetric
$(n,m)$-GNLS hierarchy using the following general rule:
\begin{eqnarray}
R_{ab} = (J_2 J^{(-1)}_1)_{ab} \equiv \Pi
\left(\begin{array}{cc}
(R_{11})_{ab}, & (R_{12})_{ab} \\
(R_{21})_{ab}, &  (R_{22})_{ab}
\end{array}\right) \Pi, \quad
\frac{\partial}{\partial t_{p+1}}
\left(\begin{array}{cc} F_a\\{\overline F_a} \end{array}\right) =
R_{ab} \frac{\partial}{\partial t_p}
\left(\begin{array}{cc} F_b\\{\overline F_b} \end{array}\right).
\label{recop0}
\end{eqnarray}
It is defined up to an arbitrary additive operator which annihilates the
column on the r.h.s. of the second relation (\ref{recop0}) and can be
represented as $C{\overline {\Pi}}$, where $C$ is an arbitrary
matrix-valued pseudo-differential operator. Substituting eqs. \p{starj}
and \p{hamstr2} into \p{recop0}, one can easily obtain the explicit
expression for $R_{ab}$,
\begin{eqnarray}
&& (R_{11})_{ab}=({\partial}+
\frac{1}{2}F_cD\overline D \partial^{-1} \overline F_c){\delta}_{ab} -
{1\over 2}(-1)^{d_b}( F_aD\overline D \partial^{-1} \overline F_b +
{\partial} F_a\partial^{-1}\overline F_b), \nonumber\\
&& (R_{12})_{ab}= \frac{1}{2}((-1)^{d_ad_b} F_bD\overline
D \partial^{-1} F_a- F_aD\overline D \partial^{-1} F_b -
{\partial} F_a{\partial}^{-1} F_b), \nonumber\\
&& (R_{21})_{ab}=\frac{1}{2}(-1)^{d_b}
((-1)^{d_ad_b}{\overline F}_b~{\overline D} D \partial^{-1} \overline F_a-
{\overline F}_a~{\overline D} D \partial^{-1} \overline F_b-
{\partial}\overline F_a\partial^{-1}\overline F_b), \nonumber\\
&& (R_{22})_{ab}=(-{\partial}+\frac{1}{2}(-1)^{d_c}{\overline F}_c~
{\overline D} D \partial^{-1} F_c){\delta}_{ab}-
\frac{1}{2}( {\overline F}_a~{\overline D} D \partial^{-1} F_b +
{\partial} \overline F_a \partial^{-1} F_b),
\label{recop}
\end{eqnarray}
and the recurrence relations for the flows,
\begin{eqnarray}
&& {\textstyle{\partial\over\partial t_{p+1}}}F_a=
{\textstyle{\partial\over\partial t_{p}}}F_a~'-
\frac{1}{2}(F_a D \overline D - D \overline D F_a){\partial}^{-1}
{\textstyle{\partial\over\partial t_{p}}}(F_b {\overline F}_b)+
\frac{1}{2} F_b D \overline D {\partial}^{-1}
{\textstyle{\partial\over\partial t_{p}}}({\overline F}_bF_a),\nonumber\\
&& {\textstyle{\partial\over\partial t_{p+1}}}{\overline F}_a=-
{\textstyle{\partial\over\partial t_{p}}}{\overline F}_a~'-
\frac{1}{2}({\overline F}_a \overline D D - \overline D D
{\overline F}_a){\partial}^{-1} {\textstyle{\partial\over\partial t_{p}}}
(F_b {\overline F}_b)+ \frac{(-1)^{d_b}}{2}{\overline F}_b \overline D D
{\partial}^{-1} {\textstyle{\partial\over\partial t_{p}}}
(F_b {\overline F}_a).~~~~~~~
\label{recrel}
\end{eqnarray}

At this point let us make a remark which will be useful in the following.
Taking into account the local nature of flows \p{n1} of the $N=2$
supersymmetric $(n,m)$-GNLS hierarchy, a simple inspection of the recurrence
relations \p{recrel} allows one to conclude that the time derivatives of
the superfunctions
\begin{eqnarray}
{\cal H}_{1,ab} \equiv F_a
{\overline F}_b
\label{nres0}
\end{eqnarray}
should be represented as the
sum of total bosonic and fermionic derivatives of some local superfield
functions. Moreover, the evolution equations for the function
${\cal H}_{1,aa}$ should contain only a total bosonic derivative. In other words
the quantities
\begin{eqnarray}
H_{1,ab} = \int d Z {\cal H}_{1,ab}, \quad
\widetilde{H}_0 = \int d z {\cal H}_{1,aa}
\label{nres}
\end{eqnarray}
are to be integrals of the flows. For the flows \p{gnls}, this can be
checked by simple direct calculations. For the $p-$th flow of the $N=2$
supersymmetric $(n,m)$-GNLS hierarchy, the evolution equation for
the ${\cal H}_{1,aa}$ takes the following general form \cite{s1}:
\begin{eqnarray}
-\frac{1}{2}{\textstyle{\partial\over\partial t_p}}
{\cal H}_{1,aa} = ((L^p)_0)~'
\label{n}
\end{eqnarray}
which agrees with the above-mentioned arguments.

Using the Poisson brackets algebra \p{palg}, \p{hamstr2} one can calculate the
Poisson brackets between the integrals $H_{1,ab}$ and ${\widetilde H}_0$ 
\p{nres}
\begin{eqnarray}
\{H_{1,ab},{\widetilde H}_0\}=0, \quad
\{H_{1,ab}, H_{1,cd}\} = 2 (-1)^{d_ad_d}{\delta}_{bc} H_{2,da}
- 2 (-1)^{d_d(d_c+d_b+1)}{\delta}_{ad} H_{2,bc},
\label{int}
\end{eqnarray}
where the new nonlocal integrals
\begin{eqnarray}
H_{2,ab} = \int d Z {\overline F}_a L F_b,
\label{nint}
\end{eqnarray}
have been introduced, $L$ being the Lax operator \p{suplax}. 
The integrals complex-conjugate
with respect to complex structure \p{conj}
$H^{*}_{2,ad}$ are related to $H_{2,da}$ as
$H^{*}_{2,ad}=(-1)^{\frac{(d_a+d_d)^{2}}{2}} H_{2,da}$. Repeatedly applying
the same procedure one can generate new series of nonlocal integrals.

Acting $p$-times with the recursion operator \p{recop0}, \p{recop} on the
zeroth flow from the set \p{gnls} and on the second Hamiltonian structure
\p{hamstr2} of the $N=2$ supersymmetric $(n,m)$-GNLS hierarchy, one can
derive $p$-th flow, as well as the $(p+2)$-th Hamiltonian structure,
\begin{eqnarray}
{\textstyle{\partial\over\partial t_p}}
\left(\begin{array}{cc} F_a\\{\overline F_a} \end{array}\right) =
(R^{p})_{ab}
\left(\begin{array}{cc} F_b\\-{\overline F_b} \end{array}\right), \quad
J_{p+2} = R^p J_2,
\label{hamstrn}
\end{eqnarray}
respectively. Substituting the explicit expressions \p{recop0}, \p{recop}
for the recursion operator into the first formula of eqs. \p{hamstrn}, we
obtain, for example, the following set of equations for the 3-th flow:
\begin{eqnarray}
&& {\textstyle{\partial\over\partial t_3}} F_a= F_a~''' + \frac{3}{2} D
({\overline D}(F_b {\overline F}_b F_a~') - \frac{1}{2} 
(F_b {\overline F}_b)^2 {\overline D} F_a +
\left[D {\overline F}_b \right] \left[{\overline D} F_b\right]
{\overline D} F_a), \nonumber\\
&& {\textstyle{\partial\over\partial t_3}} {\overline F}_a =
{\overline F}_a~''' - \frac{3}{2} {\overline D}
(D ( F_b {\overline F}_b {\overline F}_a~') + \frac{1}{2}
(F_b {\overline F}_b)^2 D {\overline F}_a - 
\left[D {\overline F}_b\right] \left[{\overline D} F_b\right]
D{\overline F}_a),
\label{threef}
\end{eqnarray} 
which coincides with the corresponding set that can be derived using eqs.
\p{n1} and gives a confirmation of the above-constructed formulas.
The flows allow the Hamiltonian densities \p{res}
corresponding to them to be constructed using eq. \p{n},
\begin{eqnarray}
{\cal H}_p=
-\frac{1}{2} {\partial}^{-1} {\textstyle{\partial\over\partial t_{p}}}
(F_b {\overline F}_b),
\label{ham}
\end{eqnarray}
without knowing the Lax operator. Thus, almost all information about the
$N=2$ supersymmetric $(n,m)$-GNLS hierarchy is encoded in its recursion
operator. 

For the particular cases $n=0, m=1$ and $n=1, m=0$, the expressions \p{starj},
\p{hamstr2} and \p{recop0}-\p{recop} for the Hamiltonian structures and the
recursion operator of the $N=2$ supersymmetric $(n,m)$-GNLS hierarchy
reproduce the corresponding expressions constructed in \cite{bks,dls}. 

{}~

\noindent{\bf 3. Hamiltonian structure of the $N=2$ super
$(n,m)$-GNLS hierarchy in the KdV-basis.}
In the KdV-basis \p{newbasis}, the general set of bi-Hamiltonian equations
\p{hameq} takes the form:
\begin{eqnarray}
&& {\textstyle{\partial\over\partial t_p}}
\left(\begin{array}{cc} J\\ {\overline \Phi}_i \\ \Phi_i
\end{array}\right) =
(J^{KdV}_1)_{ij}\left(\begin{array}{cc}
{\delta}/{\delta J} \\
{\delta}/{\delta {\overline \Phi}_j}\\
{\delta}/{\delta \Phi_j} \\
\end{array}\right) H_{p+1}=
(J^{KdV}_2)_{ij}\left(\begin{array}{cc}
{\delta}/{\delta J} \\
{\delta}/{\delta {\overline \Phi}_j}\\
{\delta}/{\delta \Phi_j} \\
\end{array}\right) H_{p}, \nonumber\\
&& (J^{KdV}_1)^{(-1)}_{ij} {\textstyle{\partial\over\partial t_p}}
\left(\begin{array}{cc}
J\\ {\overline \Phi}_j \\ \Phi_j \end{array}\right) =
\left(\begin{array}{cc}
{\delta}/{\delta J} \\
{\delta}/{\delta {\overline \Phi}_j} \\
{\delta}/{\delta \Phi_j} \\
\end{array}\right) H_{p+1}.
\label{nhameq2}
\end{eqnarray}
The Hamiltonian structures $J^{KdV}_p$ are related to $J_p$
\p{hamstrn} by the general rule\footnote{Let us recall the rules for the
adjoint conjugation operation $T$: $D^{T}=-D$, ${\overline D}^{T}=
-{\overline D}$, $(QP)^{T}=(-1)^{d_Qd_P}P^{T}Q^{T}$, where $Q$ and
$P$ are arbitrary operators. In addition, for matrices, it is necessary to
take the operation of the matrix transposition. All other rules can be
derived using these.}
\begin{eqnarray}
(J^{KdV}_p)_{ij}= {\cal G}_{ia} (J_p)_{ab} ({\cal G}^T)_{bj}, \quad
(J^{(-1)}_p)_{ab}= ({\cal G}^{T})_{ai} (J^{KdV}_p)^{(-1)}_{ij} {\cal
G}_{jb},
\label{tran}
\end{eqnarray}
where\footnote{Let us remember that the index $l$ is an arbitrary fixed
index belonging to the range $1\leq l \leq n$. Therefore, in \p{frech},
there is no summation over repeated indices $l$.}
\begin{eqnarray}
{\cal G}_{ia} \equiv \left(\begin{array}{cccc}
\frac{1}{2}(\frac{1}{2}(-1)^{d_a}{\overline F}_a -
\partial F_l^{-1}{\delta}_{al}), & \frac{1}{4} F_a \\
\frac{1}{\sqrt{2}}\overline D F_l^{-1}({\delta}_{ia} -
F_l^{-1} F_i {\delta}_{al}), & 0 \\
\frac{1}{\sqrt{2}}D{\overline F}_i {\delta}_{al}, &
\frac{1}{\sqrt{2}} D F_l {\delta}_{ia}
\end{array}\right) \Pi
\label{frech}
\end{eqnarray}
is the matrix of Fr\'echet derivatives corresponding to the transformation
$\{J,{\overline \Phi}_i,\Phi_i\}$ $\Rightarrow$ $\{F_a,$$ {\overline F}_a\}$
\p{newbasis} to the KdV-basis. Using eqs. \p{starj}, \p{hamstr2} and
\p{tran}, one can derive the following expressions for the
first\footnote{Here, $d_i$ is the Grassman parity of the superfields
$\Phi_i$ and ${\overline \Phi}_i$.},
\begin{eqnarray}
&& (J^{KdV}_1)^{(-1)}_{ij} = \left(\begin{array}{ccc}
4,  & 0, & 0 \\
0, & 0, & (-1)^{d_i} D \overline D {\partial}^{-1} {\delta}_{ij} \\
0, & \overline D D {\partial}^{-1} {\delta}_{ij}, & 0
\end{array}\right){\partial}^{-1}, \nonumber\\
&& (J^{KdV}_1)_{ij} = \left(\begin{array}{ccc}
\frac{1}{4}, & 0, & 0 \\
0, & 0, & \overline D D {\partial}^{-1} {\delta}_{ij} \\
0, & (-1)^{d_i} D \overline D {\partial}^{-1} {\delta}_{ij}, & 0
\end{array}\right)\partial, \nonumber\\
&& J^{KdV}_1 (J^{KdV}_1)^{(-1)} =
\left(\begin{array}{cc}
1, & 0 \\
0, &  {\overline \Pi}
\end{array}\right), \quad
(J^{KdV}_1)^{(-1)} J^{KdV}_1 =
\left(\begin{array}{cc}
1, & 0 \\
0, &  \Pi
\end{array}\right),
\label{j1}
\end{eqnarray}
and for the second,
\begin{eqnarray}
&& (J^{KdV}_2)_{ij} =
\left(\begin{array}{ccc}
J_{11}, & (J_{12})_{j}, & (J_{13})_{j} \\
(J_{21})_{i}, &  (J_{22})_{ij}, & (J_{23})_{ij} \\
(J_{31})_{i}, &  (J_{32})_{ij}, & (J_{33})_{ij} \\
\end{array}\right), \nonumber\\
&& J_{11}=-\frac{1}{2}(\frac{1}{2}[D, \overline D ~]\partial +\overline D
JD +DJ\overline D+\partial J +J\partial), \nonumber\\
&&(J_{12})_{j}=\frac{1}{2}({\overline \Phi}_jD+
(-1)^{d_j}D{\overline \Phi}_j)\overline D, \quad
(J_{13})_{j}=\frac{1}{2}(\Phi_j\overline D+(-1)^{d_j}\overline D \Phi_j) D,
\nonumber\\
&&(J_{21})_{i}=\frac{1}{2}\overline D((-1)^{d_i}{\overline \Phi}_iD+
D{\overline \Phi}_i), \quad
(J_{31})_{i}=\frac{1}{2}D((-1)^{d_i}\Phi_i\overline D+\overline D \Phi_i),
\nonumber\\
&& (J_{22})_{ij}={\overline \Phi}_i D\overline D{\partial}^{-1}
{\overline \Phi}_j-(-1)^{d_id_j}{\overline \Phi}_j D\overline D
{\partial}^{-1}{\overline \Phi}_i, \nonumber\\
&& (J_{33})_{ij}= -\Phi_i \overline D D{\partial}^{-1} {\overline \Phi}_j+
(-1)^{d_id_j}\Phi_j \overline D D{\partial}^{-1} \Phi_i, \nonumber\\
&& (J_{23})_{ij}=(\overline D (\partial - 2J) D -
(-1)^{d_m}{\overline \Phi}_m \overline D D{\partial}^{-1} \Phi_m)
{\delta}_{ij}  +  {\overline \Phi}_i \overline D D{\partial}^{-1} \Phi_j,
\nonumber\\
&& (J_{32})_{ij}=-(-1)^{d_j}
( D (\partial + 2J)\overline D -
\Phi_m D \overline D {\partial}^{-1} {\overline \Phi}_m)
{\delta}_{ij}  -  \Phi_i D \overline D {\partial}^{-1} {\overline \Phi}_j,
\label{j2}
\end{eqnarray}
Hamiltonian structures, respectively, and construct the recursion operator:
\begin{eqnarray}
&& R^{KdV}_{ij} = (J^{KdV}_2 (J^{KdV}_1)^{(-1)})_{ij} \equiv
\left(\begin{array}{ccc}
4J_{11}, & -(J_{13})_{j}, & -(-1)^{d_j}(J_{12})_{j} \\
4(J_{21})_{i}, & -(J_{23})_{ij}, & -(-1)^{d_j}(J_{22})_{ij} \\
4(J_{31})_{i}, & -(J_{33})_{ij}, & -(-1)^{d_j} (J_{32})_{ij}
\end{array}\right) {\partial}^{-1}, \nonumber\\
&& {\textstyle{\partial\over\partial t_{p+1}}}
\left(\begin{array}{cc} J\\ {\overline \Phi}_i \\ \Phi_i
\end{array}\right) = R^{KdV}_{ij}
{\textstyle{\partial\over\partial t_{p}}}
\left(\begin{array}{cc} J\\ {\overline \Phi}_j \\ \Phi_j
\end{array}\right)= (R^{KdV})^{p}_{ij}
\left(\begin{array}{cc} J~'\\ {\overline \Phi}_j~' \\ \Phi_j~'
\end{array}\right)
\label{nrecop}
\end{eqnarray}
of the $N=2$ supersymmetric $(n,m)$-GNLS hierarchy in the KdV-basis.

The Jacobi identities for the first Hamiltonian structure $J^{KdV}_1$
\p{j1} are obviously satisfied as for the constant-coefficient operator
with the correct symmetry properties. For the particular cases $n=1, m=0$
($n=0,m=1$) and $n=1, m=1$, $J^{KdV}_1$ was found in \cite{k,lm} and
\cite{dik}, respectively, and the hereditary recursion operator for the
former case was constructed in \cite{op}.

In regard to the Jacobi identities
for the second Hamiltonian structure \p{j2}, for the particular case
$n=1, m=0$ ($n=0, m=1$), $J^{KdV}_2$ coincides with the $N=2$
superconformal algebra which is the second Hamiltonian structure of the
$N=2$ $a=4$ KdV hierarchy \cite{lm}, and for the case $n=1,m=1$, it forms
the $N=4$ $SU(2)$ superconformal algebra---the second Hamiltonian
structure of the $N=4$ $SU(2)$-KdV hierarchy \cite{dik}. Therefore, 
for these cases, they are satisfied. We did not check them for the other values 
of
the discrete parameters $n$ and $m$. However, we have verified the
$J^{KdV}_2$ for the first four flows of the $N=2$ supersymmetric $(n,m)$-GNLS
hierarchy at arbitraty values of $n$ and $m$. Moreover, in what follows,
we check that in the bosonic limit it correctly reproduces the second
Hamiltonian structure of the bosonic GNLS hierarchy for arbitrary value
of the parameter $m$. Taking into account these arguments, it is natural to
expect that the expressions \p{j2} for the general
supersymmetric case are correct, but we do not present a proof here. 

For arbitrary values of the parameters $n$ and $m$,
the Hamiltonian structures $J^{KdV}_1$ and $J^{KdV}_2$ are obviously
compatible: the deformation of the superfield $J$
$\Rightarrow$  $J + \gamma$, where $\gamma$ is an arbitrary parameter,
transforms $J^{KdV}_2$ into the Hamiltonian structure defined by their
algebraic sum $J^{KdV}_2 - 2 {\gamma} J^{KdV}_1$. Thus, one can conclude
that the recursion operator $R^{KdV}_{ij}$ \p{nrecop} is hereditary as 
the operator obtained from the compatible pair of the Hamiltonian structures
\cite{ff}. 

Let us remark that the second Hamiltonian structures 
$J^{KdV}_2$ form the extended $N=2$ superconformal
algebras, possessing a manifest $N=2$ supersymmetry, with the $N=2$
stress-tensor $J(Z)$ and spin-1 primary fermionic and bosonic supercurrents,
$\Phi_i(Z)$ and ${\overline {\Phi}}_i(Z)$. For general values of
the parameters $n$ and $m$, these algebras are nonlocal. 
Taking into account that the $N=2$ superconformal algebra can be derived via
the Hamiltonian reduction of the $N=2$ $sl(2|1)$ affine superalgebra,
it is reasonable to conjecture the existence of a 
similar relation of our superalgebras to the $N=2$
$sl(k|k-1)$ affine superalgebras \cite{ais}. The detailed analysis of 
this complicated problem is however out the scope of the present letter.

In the KdV-basis \p{newbasis}, there are also series of
local and nonlocal additional integrals,
\begin{eqnarray}
&& H^{KdV}_0= \int d z J, \quad {\widetilde H}^{KdV}_{1,i}=
\int d z {\overline D} {\Phi}_i, \nonumber\\
&& H^{KdV}_{1,ij} =\int d Z
{\Phi}_i{\partial}^{-1}{\overline {\Phi}}_j, \quad
H^{KdV}_{1,i} =\int d Z
\left[ J -\frac{1}{2}{\Phi}_j \partial^{-1} {\overline {\Phi}}_j\right]
D \partial^{-1} {\overline {\Phi}}_i, \nonumber\\
&& {\widetilde H}^{KdV}_{2}=\int d Z \left[ J - 
\frac{1}{2} {\Phi}_j {\partial}^{-1} {\overline {\Phi}}_j \right]
L^{KdV} 1, \quad {\widetilde H}^{KdV}_{2,i}= \int d z 
\left[{\overline D} {\Phi}_i \right] L^{KdV} 1, \nonumber\\
&& H^{KdV}_{2,ij} =
\int d Z {\Phi}_i L^{KdV}{\partial}^{-1}{\overline {\Phi}}_j, \quad
H^{KdV}_{2,i} =\int d Z
\left[ J -\frac{1}{2}{\Phi}_j \partial^{-1} {\overline {\Phi}}_j\right]
L^{KdV}D \partial^{-1} {\overline {\Phi}}_i,
\label{kdvnint}
\end{eqnarray}
corresponding to the integrals \p{nres0}, \p{nres} and \p{nint}, as well as
their complex-conjugates with respect to complex structure \p{nnconj}.
Up to normalization constants, here are some of them: 
\begin{eqnarray}
&& {\widetilde H}^{*KdV}_{1,i}=\int d z D {\overline {\Phi}}_i, \quad
H^{*KdV}_{1,i}=\int d Z \left[ J + \frac{(-1)^{d_j}}{2} {\overline {\Phi}}_j
{\partial}^{-1} {\Phi}_j\right] {\overline D} {\partial}^{-1}
{\Phi}_i, \nonumber\\ 
&& {\widetilde H}^{*KdV}_{2,i}=\int d z \left[D {\overline {\Phi}}_i\right]
L^{*KdV} 1, \quad H^{*KdV}_{2,i}= \int d Z
\left[ J+ \frac{(-1)^{d_j}}{2} {\overline {\Phi}}_j {\partial}^{-1} {\Phi}_j
\right] L^{*KdV} {\overline D} {\partial}^{-1} {\Phi}_i. ~ ~ ~ ~ 
\label{ckdvint}
\end{eqnarray}
These are algebraically independent with respect to integrals
\p{kdvnint}. Here, $L^{KdV}$ is the Lax operator \p{kdvsuplax}, and
$L^{*KdV}$ is its complex-conjugate operator with respect to complex
structure \p{nnconj}.

Let us remark that for the particular case $n=1, m=1$, the Poisson
barckets between the superfield integrals $H^{KdV}_0$, ${\widetilde
H}^{KdV}_{1,1}$ and ${\widetilde H}^{*KdV}_{1,1}$, calculated using the
second Hamiltonian structure $J^{KdV}_2$, form the global $N=4$ supersymmetric
algebra in one dimension. The Poisson brackets between these integrals and
the superfields $J$, $\Phi_1$ and ${\overline \Phi}_1$ generate the
$N=4$ infinitesimal transformations of the last ones, which are symmetry
transformations of the $N=4$ $SU(2)$-KdV hierarchy. As an example, 
we present the transformations generated by the sum 
${\overline {\epsilon}}{\widetilde H}^{KdV}_{1,1} + {\epsilon}{\widetilde
H}^{*KdV}_{1,1}$ of the integrals,
\begin{eqnarray}
{\delta} J=\frac{1}{2}({\epsilon}D {\overline {\Phi}}_1 + {\overline
{\epsilon}}~{\overline D} {\Phi}_1), \quad {\delta} {\Phi}_1=-2 {\epsilon} 
D J, \quad {\epsilon}{\overline {\Phi}}_1= -2 {\overline {\epsilon}}~
{\overline D} J,  
\label{t1}
\end{eqnarray}
which coincide with the transformations of the hidden
$N=2$ supersymmetry of the $N=4$ $SU(2)$-KdV hierarchy, derived in \cite{dik}.
Here, $\epsilon$ and ${\overline {\epsilon}}$ are the fermionic parameters
of the transformation. In the former superfield basis $\{F_a, {\overline
F}_a\}$, the $N=4$ supersymmetric transformations are generated by the
integrals ${\widetilde H}_0$ and $H_{1,12}$
\p{nres0}, \p{nres} as well as  by the integral 
\begin{eqnarray}
{\widetilde H}^{*}_{1,1}=\int d Z \frac{F_2}{F_1},
\label{yyy}
\end{eqnarray}
which corresponds to the integral ${\widetilde H}^{*KdV}_{1,1}$. In this basis
the transformations \p{t1} become\footnote{To derive the transformations
generated by integral \p{yyy}, it is necessary to remove the ambiguity in
the operators ${\overline D}D{\partial}^{-1}1$ and
${\partial}{\partial}^{-1}1$
that appear in the calculations by setting ${\overline D}D{\partial}^{-1}1=
({\overline D}D{\partial}^{-1})1 \equiv -1$ and 
${\partial}{\partial}^{-1}1=({\partial}{\partial}^{-1})1 \equiv 1$. 
Let us remark that in spite of the chiral nature of the integrated function
$\frac{F_2}{F_1}$, in general, the integral \p{yyy} is not equal to zero
due to its singularity, and the surface terms should be taken into account.} 
\begin{eqnarray}
&& {\delta} F_a= \frac{(-1)^{d_a}}{\sqrt{2}} ({\overline {\epsilon}}
((2{\partial}+F_bD{\overline D}{\partial}^{-1}{\overline F}_b)
F_1{\delta}_{a2}+F_1D{\overline D}{\partial}^{-1}F_a{\overline F}_2) +
{\epsilon}F_2 {\delta}_{a1}), \nonumber\\
&& {\delta} {\overline F}_a=  \frac{(-1)^{d_a}}{\sqrt{2}}
({\overline {\epsilon}}((-2{\partial}+ (-1)^{d_b}{\overline F}_b
{\overline D} D {\partial}^{-1}F_b) {\overline F}_2 {\delta}_{a1}-
 (-1)^{d_a}{\overline F}_2 ~{\overline D}D{\partial}^{-1}{\overline F}_aF_1)-
{\epsilon}{\overline F}_1{\delta}_{a2}).~ ~ ~ ~ ~ ~ ~
\label{t2}
\end{eqnarray}
We have checked that transformations \p{t2} are indeed the symmetry
transformations for the $N=2$ supersymmetric $(1,1)$-GNLS equations
\p{gnls} and that their Lie brackets coincide
with the brackets for the transformations \p{t1}. Thus, the $N=2$
supersymmetric $(1,1)$-GNLS hierarchy can also be called the $N=4$
supersymmetric NLS-mNLS hierarchy reflecting the name of its first
nontrivial bosonic representative (see the next section)\footnote{For
other examples of $N=4$ supersymmetric NLS-type integrable
hierarchies, see the recent paper \cite{ikt}}. In fact, it possesses
one more $N=4$ supersymmetry which one can derive by the 
complex-conjugation operation with respect to the first
complex structure \p{conj} applied either to the integrals ${\widetilde H}_0$,
$H_{1,12}$ and ${\widetilde H}^{*}_{1,1}$ or directly to their $N=4$
supersymmetry transformations. One can easily observe that under this
operation the first integral, which form the standard $N=2$ supersymmetry,
only change the overall sign, while the other two integrals are drastically
changed (together with the transformations generated by them). Thus, one can
conclude that these two different $N=4$ supersymmetries intersect along 
the $N=2$ supersymmetry. Without going to more details, we present the new
transformations, 
\begin{eqnarray}
&& {\delta}F_a =\frac{(-1)^{d_a}}{\sqrt{2}}({\widetilde
{\epsilon}}((2{\partial}+
F_bD{\overline D}{\partial}^{-1}{\overline F}_b)F_2{\delta}_{a1}-
(-1)^{d_a}F_2D{\overline D}{\partial}^{-1}F_a{\overline F}_1)-
{\overline {\widetilde {\epsilon}}}F_1{\delta}_{a2}), \nonumber\\
&& {\delta}{\overline F}_a=\frac{(-1)^{d_a}}{\sqrt{2}}({\widetilde {\epsilon}}
((2 {\partial}-(-1)^{d_a}{\overline F}_b{\overline D}D{\partial}^{-1}F_b)
{\overline F}_1{\delta}_{a2}+{\overline F}_1{\overline D}D{\partial}^{-1}
{\overline F}_aF_2)- {\overline {\widetilde {\epsilon}}}~{\overline F}_2
{\delta}_{a1}),
\label{t3}
\end{eqnarray}  
which are the counterparts of the transformations \p{t2}. Here, 
${\widetilde {\epsilon}}$ and ${\overline {\widetilde {\epsilon}}}$ are 
the two new independent fermionic parameters. We have also checked
that the algebraic closure of these two sets of $N=4$ supersymmetric 
generator integrals contain new integrals, but the detailed analysis
of the resulting algebra will be discussed elsewhere. Let us
only mention that the transformations with the parameters $\epsilon$ and 
${\overline {\widetilde {\epsilon}}}$ in the closing generate the
transformations of the $GL(1|1)$ supergrop\footnote{Let us recall
that, for general values of the parameters $n$ and $m$, the $N=2$ supersymmetric
$(n,m)$-GNLS hierarchy is invariant with respect to $GL(n|m)$ supergrop
\cite{bks}.}.  

As for generic values of the parameters $n$ and $m$, the algebras of
the corresponding integrals and the transformation properties of the
superfields can be derived in a similar way, but again their detailed 
description will not be given here.

{}~

\noindent{\bf 4. Bosonic limit of the $N=2$ super $(n,m)$-GNLS
Hamiltonian structure.}
To derive the bosonic limit, we set all fermionic components of the
superfields $F_a$ and $\overline F_a$ equal to zero and define the bosonic
components as \cite{bks}
\begin{eqnarray}
&& b_{\alpha}  = \frac{1}{\sqrt{2}} F_{\alpha}|, \quad
{\overline b}_{\beta} = \frac{1}{\sqrt{2}} {\overline F}_{\beta}|, \quad
1 \leq {\alpha},{\beta} \leq n, \nonumber\\
&& g_s = \frac{1}{\sqrt{2}} {\overline D} F_{s+n}|~
{\exp (-{\partial^{-1}} (b_{\beta} {\overline b}_{\beta}))}, \quad
{\overline g}_p = \frac{1}{\sqrt{2}} D {\overline F}_{p+n}|~
{\exp ({\partial^{-1}} (b_{\beta} {\overline b}_{\beta}))}, \quad
1 \leq s,p \leq m, ~ ~ ~ ~ ~
\label{def}
\end{eqnarray}
where $|$ means the $({\theta}, {\bar\theta})\rightarrow 0$ limit.
In terms of such components, the second flow equations \p{gnls} for the
fields $b_{\alpha},{\overline b}_{\alpha}$ and $g_s, \overline g_s$ are
completely decoupled:
\begin{eqnarray}
{\textstyle{\partial\over\partial t_2}} b_{\alpha}=
b_{\alpha}~'' -  2 b_{\beta} {\overline b}_{\beta} b_{\alpha}~', \quad
{\textstyle{\partial\over \partial t_2}} {\overline b}_{\alpha} =
-{\overline b}_{\alpha}~'' - 2 b_{\beta}
   {\overline b}_{\beta} {\overline b}_{\alpha}~',
\label{bgnls}
\end{eqnarray}
\begin{eqnarray}
{\textstyle{\partial\over\partial t_2}} g_s=
g_s~'' -  2 g_p {\overline g}_p g_s, \quad
{\textstyle{\partial\over \partial t_2}} {\overline g}_s =
-{\overline g}_s~'' + 2 g_p {\overline g}_p {\overline g}_s.
\label{fgnls}
\end{eqnarray}
The set of equations \p{fgnls} and \p{bgnls} form the bosonic GNLS
\cite{fk} and modified GNLS (mGNLS) \cite{bks} equations, respectively.

The bosonic limit of the Hamiltonian structures \p{starj} and \p{hamstr2},
recursion operator \p{recop0}, \p{recop}, and integrals \p{nres0}, \p{nres}
and \p{nint}, corresponding to equations \p{bgnls} and \p{fgnls}, also
splits into two independent structures, which one can see from the
following explicit expressions:
\begin{eqnarray}
(J^{mGNLS}_1)^{-1}_{{\alpha}{\beta}}=
\left(\begin{array}{cc}
{\overline b}_{\alpha}~' \partial^{-1} {\overline b}_{\beta} +
{\overline b}_{\alpha} \partial^{-1} {\overline b}_{\beta}~',
& {\overline b}_{\alpha}~' \partial^{-1} b_{\beta} +
{\overline b}_{\alpha} \partial^{-1} b_{\beta} \partial+
\partial{\delta}_{{\alpha}{\beta}}\\
-b_{\alpha}~' \partial^{-1} {\overline b}_{\beta} -
b_{\alpha} \partial^{-1} {\overline b}_{\beta} \partial +
\partial{\delta}_{{\alpha}{\beta}},
& -b_{\alpha}~' \partial^{-1} b_{\beta}-b_{\alpha} \partial^{-1}b_{\beta}~'
\end{array}\right),
\label{bstarj}
\end{eqnarray}
\begin{eqnarray}
(J^{GNLS}_1)^{-1}_{sp}=
\left(\begin{array}{cc}
       0,       & -{\delta}_{sp}  \\
{\delta}_{sp}, &  0
\end{array}\right), \quad
(J^{GNLS}_1)_{sp}=
\left(\begin{array}{cc}
       0,       & {\delta}_{sp}  \\
-{\delta}_{sp}, &  0
\end{array}\right),
\label{fstarj}
\end{eqnarray}
for the first, and,
\begin{eqnarray}
(J^{mGNLS}_2)_{{\alpha}{\beta}}=
\left(\begin{array}{cc}
b_{\beta}  \partial^{-1} b_{\alpha} -
b_{\alpha} \partial^{-1} b_{\beta}, &
(1 -b_{\gamma}\partial^{-1}{\overline b}_{\gamma})
{\delta}_{{\alpha}{\beta}}+
b_{\alpha}\partial^{-1}{\overline b}_{\beta}, \\
-(1 +{\overline b}_{\gamma}\partial^{-1}b_{\gamma})
{\delta}_{{\alpha}{\beta}}+
{\overline b}_{\alpha}\partial^{-1}b_{\beta}, &
{\overline b}_{\beta} \partial^{-1} {\overline b}_{\alpha}-
{\overline b}_{\alpha} \partial^{-1} {\overline b}_{\beta},
\end{array}\right),
\label{bhamstr2}
\end{eqnarray}
\begin{eqnarray}
(J^{GNLS}_2)_{sp}=
\left(\begin{array}{cc}
g_p \partial^{-1} g_s + g_s \partial^{-1} g_p, &
(\partial - g_c\partial^{-1}{\overline g}_c)
{\delta}_{sp} - g_s\partial^{-1}{\overline g}_p, \\
(\partial - {\overline g}_c\partial^{-1} g_c)
{\delta}_{sp} - {\overline g}_s\partial^{-1} g_p, &
{\overline g}_p \partial^{-1} {\overline g}_s+
{\overline g}_s \partial^{-1} {\overline g}_p,
\end{array}\right),
\label{fhamstr2}
\end{eqnarray}
for the second Hamiltonian structures, and,
\begin{eqnarray}
R^{mGNLS}_{{\alpha}{\beta}} = \left(\begin{array}{cc}
(1 - b_{\gamma}\partial^{-1} {\overline b}_{\gamma}){\partial}
{\delta}_{{\alpha}{\beta}} -
b_{\alpha}~'\partial^{-1} {\overline b}_{\beta}, &
b_{[{\alpha},}\partial^{-1} b_{{\beta}]}+
[b_{\alpha}\partial^{-1} b_{\beta},\partial ] \\
{\overline b}_{[{\alpha},}\partial^{-1} {\overline b}_{{\beta}]}+
[~{\overline b}_{\alpha}\partial^{-1} {\overline b}_{\beta},\partial ],&
-(1 + {\overline b}_{\gamma}\partial^{-1} b_{\gamma}){\partial}
{\delta}_{{\alpha}{\beta}} -
{\overline b} _{\alpha}~'\partial^{-1} b_{\beta}
\end{array}\right),
\label{brecop}
\end{eqnarray}
\begin{eqnarray}
R^{GNLS}_{sp}=\left(\begin{array}{cc}
(\partial - g_c\partial^{-1}{\overline g}_c)
{\delta}_{sp} - g_s\partial^{-1}{\overline g}_p, &
-g_p \partial^{-1} g_s - g_s \partial^{-1} g_p \\
{\overline g}_p \partial^{-1} {\overline g}_s+
{\overline g}_s \partial^{-1} {\overline g}_p,&
(-\partial + {\overline g}_c\partial^{-1} g_c)
{\delta}_{sp} + {\overline g}_s\partial^{-1} g_p
\end{array}\right),
\label{frecop}
\end{eqnarray}
for the recursion operator, as well as,
\begin{eqnarray}
&& H^{mGNLS}_0 = \int d z b_{\alpha} {\overline b}_{\alpha}, \quad
H^{mGNLS}_{1,{\alpha}{\beta}} =
\int d z b_{\alpha} {\overline b}_{\beta}~', \nonumber\\
&& H^{mGNLS}_{2,{\alpha}{\beta}} =
\int d z {\overline b}_{\alpha} {\partial}  
( 1 - b_{\gamma}{\partial}^{-1}{\overline b}_{\gamma}) b_{\beta}~'
\equiv \int d z {\overline b}_{\alpha} L^{mGNLS}b_{\beta}~',
\label{bosint0} 
\end{eqnarray}
\begin{eqnarray}
H^{GNLS}_{1,sp} =  \int d z g_s {\overline g}_p, \quad
H^{GNLS}_{2,sp} =
\int d z {\overline g}_s (\partial - g_c {\partial}^{-1}{\overline g}_c)
g_p \equiv \int d z {\overline g}_s L^{GNLS} g_p,
\label{bosint}
\end{eqnarray}
for the integrals, where $L^{mGNLS}$ and $L^{GNLS}$ are the Lax operators
of the mGNLS and GNLS hierarchies\footnote{The operator $L^{mGNLS}$ is 
gauge-related to the Lax operator ${\widetilde L}_2$ proposed in \cite{bks}
for the mGNLS hierarchy, $L^{mGNLS}= G^{-1}{\widetilde L}_2 G$, where
$G \equiv \exp (-{\partial}^{-1}(b_{\beta}{\overline b}_{\beta}))$.}. In
the bosonic limit, the expressions for the Hamiltonian densities \p{res}
given by eq. \p{ham} look as follows: 
\begin{eqnarray}
{\cal H}^{mGNLS}_p={\partial}^{-1} {\textstyle{\partial\over\partial t_p}}
(b_{\alpha} {\overline b}_{\alpha}~'),
\label{mg}
\end{eqnarray}
\begin{eqnarray}
{\cal H}^{GNLS}_p={\partial}^{-1}{\textstyle{\partial\over\partial t_p}}
(g_s {\overline g}_s).
\label{g}
\end{eqnarray}
The Hamiltonian structures \p{fstarj}, \p{fhamstr2} and \p{g}, as well
as the recursion
operator \p{frecop} of the GNLS hierarchy, reproduce the corresponding
expressions constructed in \cite{agz}. Regarding Hamiltonian
structures \p{bstarj} and \p{bhamstr2}, as well as recursion operator
\p{brecop} for the mGNLS hierarchy, they coincide for the particular case
$n=1$ with the corresponding expressions obtained in \cite{dl}. However,
for a general value of $n$, to our knowledge, they are presented for the
first time.

{}~

\noindent{\bf 5. Conclusion.}
In this Letter, we have constructed the first and second Hamiltonian
structures, \p{starj}, \p{hamstr2}, \p{j1} and \p{j2}, as well as the
recursion operators, \p{recop0}, \p{recop} and \p{nrecop}, which connect all
evolution systems and Hamiltonian structures of the $N=2$ supersymmetric
$(n,m)$-GNLS hierarchy in two different superfield bases characterized by local
evolution equations. For general values of $n$ and $m$, to our
knowledge, they are presented here for the first time. We have also produced 
their bosonic counterparts \p{bstarj}--\p{frecop}. Finally we have constructed
the new local and nonlocal bosonic and fermionic integrals \p{nres0}, 
\p{nres}, \p{nint}, \p{kdvnint}, \p{ckdvint}, \p{yyy}, \p{bosint0} 
and \p{bosint} of the supersymmetric and bosonic hierarchies.

{}~

\noindent{\bf Acknowledgments.}
A.S. would like to thank SISSA for the hospitality during the course
of this work and for financial support.
This work was partially supported by the Russian Foundation for Basic
Research, Grant No. 96-02-17634, RFBR-DFG Grant No 96-02-00180,
INTAS Grant No. 93-1038,
INTAS Grant No. 94-2317, and by a grant from the Dutch NWO organization.

\end{document}